\documentclass[prb,twocolumn,aps,amssymb,showpacs,superscriptaddress]{revtex4-1}

\usepackage{epsfig}
\usepackage{graphicx}
\usepackage{color}
\usepackage{hyperref}

\begin{document}

\title{Pairing effects in the normal phase of a two-dimensional Fermi gas}

\author{F. Marsiglio}
\affiliation{School of Science and Technology, Universit\`{a} di Camerino, 62032 Camerino (MC), Italy}
\affiliation{Department of Physics, University of Alberta, Edmonton, Alberta, Canada T6G 2G7}

\author{P. Pieri}
\affiliation{School of Science and Technology, Universit\`{a} di Camerino, 62032 Camerino (MC), Italy}

\author{A. Perali}
\affiliation{School of Pharmacy, Universit\`{a} di Camerino, 62032 Camerino (MC), Italy}

\author{F. Palestini}
\affiliation{School of Science and Technology, Universit\`{a} di Camerino, 62032 Camerino (MC), Italy}

\author{G. C. Strinati}
\affiliation{School of Science and Technology, Universit\`{a} di Camerino, 62032 Camerino (MC), Italy}
\affiliation{INFN, Sezione di Perugia, 06123 Perugia (PG), Italy} 

                
\begin{abstract}
In a recent experiment \cite{Koehl-2011,Koehl-2012}, a pairing gap was detected in a two-dimensional (2D) Fermi gas with attractive interaction at temperatures where superfluidity does not occur. The question remains open whether this gap is a pseudo-gap phenomenon or is due to a molecular state.  In this paper, by using a t-matrix approach we reproduce quite well the experimental data for a 2D Fermi gas, and set the boundary between the pseudo-gap and molecular regimes. We also show that pseudo-gap phenomena occurring in 2D and 3D can be related through a variable spanning the BCS-BEC crossover in a universal way.
\end{abstract}                

\maketitle

\section{Introduction} 
\label{sec:introduction}
\vspace{-0.2cm}

It is commonly believed that the normal phase of high-temperature (cuprate) superconductors is more intriguing than the superfluid phase below the critical temperature $T_{c}$,
owing especially to the appearance of a \emph{pseudogap} from the observation of a suppression of low-energy weight in the single-particle spectral function above $T_{c}$ 
\cite{Sharapov-2001,Larkin-2005}.
Two alternative mechanisms have been proposed in this respect, namely, the presence of a competing order which affects \emph{directly} the single-particle properties, and the occurrence of precursor pairing in the normal phase through strong pairing fluctuations which are amplified by the (quasi) 2D nature of these systems and affect the single-particle properties only \emph{indirectly} through two-particle effects \cite{Campuzano-2008,Kaminski-2009,Shen-2010,Kaminski-2011,Lanzara-2013}.

Since it is rather difficult to isolate these two mechanisms in a solid-state material, experiments have recently been performed with ultra-cold Fermi gases using momentum-resolved radio-frequency spectroscopy (in analogy to angle-resolved photoemission spectroscopy in solid state), originally in 3D \cite{Jin-2008,Jila_Camerino-2010} and more recently in 2D \cite{Koehl-2011,Koehl-2012}. 
This is because in ultra-cold gases one can tune the inter-particle attractive interaction in a controlled way, from weak coupling with largely overlapping Cooper pairs to strong coupling where composite bosons form, amplifying in this way the occurrence of pairing fluctuations and leaving aside other forms of long-range order not connected with pairing.

Above $T_{c}$, pairing correlations between opposite-spin fermions still extend over a finite distance.  
While 3D and 2D are not qualitatively different in this respect, the temperature window above $T_{c}$ in which precursor pairing occurs is expected to be wider in 2D than in 3D, making 2D systems best suited for the study of pseudo-gap phenomena in the normal phase.
The problem is, however, that in 2D a two-body bound state occurs for any value of the inter-particle coupling, so that one may not \emph{a priori} be confident that the measured pairing gap is truly a pseudo-gap due to collective (many-body) effects like the pairing gap below $T_{c}$, or is rather a molecular binding energy.
We shall address this crucial issue by performing a theoretical study of the dispersion relation associated with the single-particle spectral function for various couplings and temperatures, determining in this way the fate of the underlying Fermi surface whose presence guarantees that the pairing gap is a truly many-body effect.

This study carries along another important issue, about the use of a more fundamental variable than the coupling to follow the evolution from the weak (BCS) to the strong (BEC) coupling limits. 
This issue was raised in Ref.~\onlinecite{PS-1994} to connect the physics of the BCS-BEC crossover with high-temperature superconductors.
Also for ultra-cold gases, however, it would be preferable to deal with 3D and 2D systems on a comparable footing through the use of a unifying variable.
It will turn out that in terms of this variable (identified as the ratio of the pair size to the average inter-particle distance) the collapse of the Fermi surface occurs \emph{simultaneously} in 3D and 2D, thus establishing a coherent framework for precursor-pairing phenomena in different dimensions. 

Recent theoretical approaches that have addressed the experiment of Ref.~\onlinecite{Koehl-2011} include the work of: 
(a) Ref.~\onlinecite{Pietila-2012} which used a non-self-consistent t-matrix approach limited to a homogeneous system and where the energy distribution curves were not wave-vector resolved; 
(b) Ref.~\onlinecite{Ohashi-2013} which considered the trapped case as well, where quantitative comparison with the experimental data was rather limited; 
(c) Refs.~\onlinecite{Parish-2013} and \onlinecite{Hofmann-2013} which used a high-temperature expansion valid for temperatures much larger than the Fermi temperature and limited to a homogeneous system, where no quantitative comparison with the experimental data was attempted;
(d) Ref.~\onlinecite{Enss-2013} which used a self-consistent t-matrix approach limited to a homogeneous system and where again no quantitative comparison with the experimental data was attempted.

The key new physical results that we have obtained can be summarized as follows:

\noindent
(i) We present a \emph{direct comparison} between our calculations and the experimental spectra of Ref.~\onlinecite{Koehl-2011} for a trapped Fermi system.
The favorable agreement we obtain in this way gives us confidence that the t-matrix approach we shall adopt throughout is an appropriate theoretical tool (at least) in the temperature range where we are going to use it. 
This is intermediate between the low-temperature region when one approaches the critical temperature \cite{Jochim-2014} from above and the high-temperature region where the t-matrix approach reduces to the virial expansion \cite{Combescot-2006}. 

\noindent
(ii) We discuss in detail the \emph{distinction} between the pseudo-gap and molecular regimes for the 2D Fermi system.
Adopting a suitable criterion for this distinction is particularly relevant in 2D where a two-body bound state is present for all couplings throughout the BCS-BEC crossover.
By applying this criterion to the results of the data of Ref.~\onlinecite{Koehl-2011}, we can assess that this experiment was indeed able to explore also the pseudo-gap regime of most physical interest.

\noindent
(iii) We exploit the \emph{common features} of the above criterion between 2D and 3D, to connect the evolution from the BCS to the BEC limit which occurs in 2D with that in 3D
(previously discussed in Ref.~\onlinecite{Camerino_Jila-2011}).
This connection requires the use of a common variable that does not depend on the differences between the two-body scattering in 2D and 3D.
To this end, we identify the appropriate variable as the ratio between the pair size and the average inter-particle distance.
Remarkably, we find that the boundary between the pseudo-gap and molecular regimes in 2D and 3D occurs at a common value of this variable.

\noindent
(iv) In terms of this new variable to scan the BCS-BEC crossover, we are able to assess in a quantitative way that the \emph{pseudo-gap regime} extends to much higher temperatures in 2D than in 3D, more than doubling the value in 2D with respect to 3D in the crossover range of interest.

The paper is organized as follows.
Section II discusses the t-matrix approach that we have used in 2D and recalls features of the temperature and coupling dependence of the chemical potential.
Section III provides a detailed comparison with the experimental data of Ref.~\onlinecite{Koehl-2011} for a 2D trapped Fermi gas in the relevant temperature range.
Section IV dwells on the problem of the 2D homogeneous Fermi gas and discusses the issue of the boundary between the pseudo-gap and molecular regimes, and identifies a unifying variable to span the BCS-BEC crossover in 2D and 3D.
Section V gives our conclusions.
The Appendix focuses on the BEC limit in 2D.

\vspace{-0.3cm}
\section{The $\mathrm{t}$-matrix approach in 2D} 
\label{sec: t-matrix}

\vspace{-0.2cm}
\begin{center}
{\bf A. General formalism}
\end{center}
\vspace{-0.1cm}

Our calculations rest on a non-self-consistent t-matrix approach, which was extensively used in 3D and is extended here to 2D.
For an inter-particle interaction of the contact type, the pair propagator is given by
\begin{equation}
\Gamma_{0}(\mathbf{q},\Omega_{\nu}) = - \frac{1}{\frac{m}{2 \pi} \, \eta + R_{pp}(\mathbf{q},\Omega_{\nu})} \, .
\label{pair-propagator}
\end{equation}

\noindent
Here, $\mathbf{q}$ is a wave vector, $\Omega_{\nu}=2 \pi k_{B} T \nu$ ($\nu$ integer) a bosonic Matsubara frequency at temperature $T$ ($k_{B}$ being the Boltzmann constant), 
and $\eta = - \ln(k_{F} a_{2D})$ is the coupling parameter defined in terms of the Fermi wave vector $k_{F}$ (which is related to the density $n$ by $k_{F} = \sqrt{2 \pi n}$) and the 2D scattering length $a_{2D}$ (which is identified by the two-body binding energy $\varepsilon_{0} = 1/(m \, a_{2D}^{2})$, $m$ being the particle mass).
In the above expression, $R_{pp}$ is the regularized particle-particle bubble
\begin{small}
\begin{eqnarray}
& & \hspace{-0.5cm} R_{pp}(\mathbf{q},\Omega_{\nu}) = k_{B} T \sum_{n} \, \int \!\! \frac{d\mathbf{k}}{(2 \pi)^{2}} \, 
G_{0}(\mathbf{k} + \mathbf{q},\omega_{n} + \Omega_{\nu}) \, G_{0}(\mathbf{k},\omega_{n})
\nonumber \\
& & \hspace{1.2cm} - \, \int \!\! \frac{d\mathbf{k}}{(2 \pi)^{2}} \, \frac{1}{ \frac{\mathbf{k}^{2}}{m} + \varepsilon_{0}} \, - \, \frac{m}{2 \pi} \, \eta
\label{regularized-pp-bubble} \\
& = & \! \int \!\! \frac{d\mathbf{k}}{(2 \pi)^{2}} \! \left\{ \!\! \frac{\left[ 1 - f_{F}(\xi(\mathbf{k})) - f_{F}(\xi(\mathbf{k}+\mathbf{q})) \right ]}
                                                                             {\xi(\mathbf{k}) + \xi(\mathbf{k}+\mathbf{q}) - i \Omega_{\nu}} 
- \frac{m}{\mathbf{k}^{2}} \, \Theta(|\mathbf{k}| - k_{F}) \!\! \right\}                                                                                                                                             
\nonumber
\end{eqnarray}
\end{small}
\noindent
\hspace{-0.4cm} where $G_{0}(\mathbf{k},\omega_{n}) = \left[ i \omega_{n} - \xi(\mathbf{k}) \right]^{-1}$ is the bare single-particle Green's function with fermionic Matsubara frequency $\omega_{n}=(2n+1) \pi k_{B} T$ ($n$ integer), $\xi(\mathbf{k}) = \frac{\mathbf{k}^{2}}{2 m} - \mu$ ($\mu$ being the fermionic chemical potential), $f_{F}(\epsilon) = (\exp{(\epsilon/k_{B}T)} + 1)^{-1}$ the Fermi function, and $\Theta(x)$ is the Heaviside unit step function of argument $x$.
[Throughout, we consider a spin-balanced system (and set $\hbar = 1$).]

Within the non-self-consistent t-matrix approach \cite{PPSC-2002}, the pair propagator (\ref{pair-propagator}) enters the fermionic single-particle self-energy in the form
\begin{equation}
\Sigma(\mathbf{k},\omega_{n}) \! = \! -\!\! \int \!\! \frac{d\mathbf{q}}{(2 \pi)^{2}} \, k_{B}T \sum_{\nu} \Gamma_{0}(\mathbf{q},\Omega_{\nu}) \, 
                                                                                                                                                 G_{0}(\mathbf{q}-\mathbf{k},\Omega_{\nu}-\omega_{n}) \, .
\label{self-energy}
\end{equation}
\noindent
The self-energy (\ref{self-energy}), in turn, enters the dressed Green's function $G(\mathbf{k},\omega_{n}) = \left[ G_{0}(\mathbf{k},\omega_{n})^{-1} - \Sigma(\mathbf{k},\omega_{n}) \right]^{-1}$.
With the analytic continuation $i \omega_{n} \rightarrow \omega + i 0^{+}$ to real frequency $\omega$, one finally ends up with the retarded Green's function $G^{R}(\mathbf{k},\omega)$ from which the desired spectral function $A(\mathbf{k},\omega) = - \frac{1}{\pi} \mathrm{Im}\left\{G^{R}(\mathbf{k},\omega)\right\}$ can be calculated \cite{FW-1971}.
Knowledge of $A(\mathbf{k},\omega)$, in turn, yields the density of states
\begin{equation}
N(\omega) = 2 \! \int \!\! \frac{d\mathbf{k}}{(2 \pi)^{2}} \, A(\mathbf{k},\omega)
\label{density-of-states}
\end{equation} 

\noindent
where the factor of $2$ accounts for the spin degeneracy, as well as the equation for the density
\begin{equation}
n = \int_{-\infty}^{+\infty} \!\! d\omega \, N(\omega) \, f_{F}(\omega) 
\label{density-equation}
\end{equation}

\noindent
which determines the chemical potential in terms of $n$.

The above definition of the coupling parameter $\eta$ in 2D complies with the requirement of ranging from $- \infty$ in extreme weak coupling to $+ \infty$ in extreme strong coupling, in analogy with the coupling parameter $(k_{F} a_{3D})^{-1}$ in 3D where $ a_{3D}$ is the scattering length in 3D.
With this choice of $\eta$, the comparison between the 2D and 3D pseudo-gap physics appears to be quite natural.
Other works, which did not address this comparison, utilized a different sign choice for the coupling parameter in 2D \cite{Giorgini-2011,Pietila-2012,Parish-2013}.
 
The non-self-consistent t-matrix approach offers some advantages over the self-consistent one which has been used in the literature also in 2D:
(i) Analytic continuation to the real frequency $\omega$ is exact within the non-self-consistent approach and does not have to rely on numerical procedures;
(ii) In 2D the self-consistent t-matrix approach makes the unphysical prediction that composite bosons remain interacting even when extremely dilute.
The non-self-consistent t-matrix approach does not suffer from this drawback and can thus better describe the BEC limit of the crossover.
The comparison between these two t-matrix approaches in the BEC limit is discussed in the Appendix;
(iii) From the computational side, the non-self-consistent t-matrix approach proves far less demanding than the self-consistent one. 
Accordingly, it is more manageable to apply when an averaging over the trap is required.

The lack of self-consistency, on the other hand, yields an unphysical behavior of the chemical potential when the $T=0$ limit is approached, as it was originally discussed in 
Ref.~\onlinecite{Schmitt-Rink-1989} and is recalled below in subsection \ref{sec: t-matrix}-B.
To the extent that we are not interested in reaching such low-temperature region, the non-self-consistent t-matrix approach that we adopt here appears to be ideally suited to address the problem of the spectral function of a 2D Fermi gas with a strong pairing interaction.

\begin{center}
{\bf B. Chemical potential vs temperature for the homogeneous system}
\end{center}
\vspace{-0.1cm}

\begin{figure}[t]
\includegraphics[angle=0,width=8.5cm]{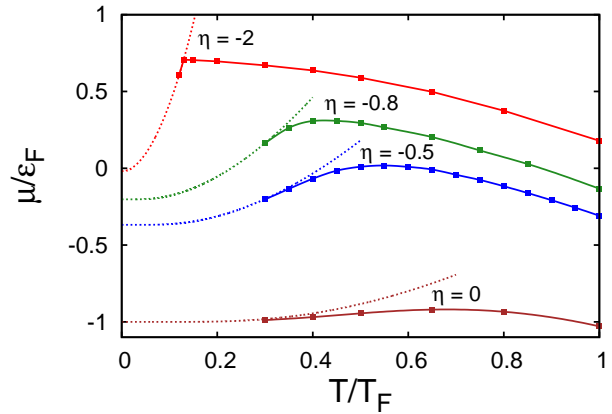}
\caption{(Color online) Chemical potential $\mu$ (dots and interpolating lines) as a function of temperature $T$  for the 2D homogeneous Fermi gas, for different values of the coupling parameter $\eta$. Dotted lines correspond to the Thouless' criterion $\Gamma_0(0,0)^{-1}=0$ calculated at the given value of $\eta$.
Here, $\varepsilon_{F} = k_{F}^{2}/(2m)$ is the Fermi energy of the homogeneous system and $T_{F} = \varepsilon_{F}/k_{B}$ the corresponding Fermi temperature.}
\label{Figure-1}
\end{figure}

Figure \ref{Figure-1} shows the chemical potential vs temperature as obtained within the present non-self-consistent t-matrix approach. 
For each coupling considered in Fig.~\ref{Figure-1}, the chemical potential presents a non-monotonic behavior reaching a maximum at an intermediate temperature. 
This behavior was already discussed some time ago by Schmitt-Rink, Varma, and Ruckenstein \cite{Schmitt-Rink-1989} (within a Nozieres-Schmitt-Rink approach \cite{NSR-1985}, which corresponds to a simplified variant of the present formalism where the Dyson's equation for the single-particle Green's function is expanded to first order in the self-energy). 
The origin of this behavior can be traced in the infrared divergence of the single-particle self-energy (\ref{self-energy}) when the Thouless' criterion curve (defined by the equation 
$\Gamma_{0}^{-1}({\bf q}=0, \Omega_{\nu}=0)=0$) is approached.  

More specifically, the equation  $\Gamma_{0}^{-1}(0, 0)=0$ defines, for each value of the coupling parameter $\eta$, a curve $\mu_{\rm c}(T)$ (dotted lines in Fig.~\ref{Figure-1}) that cannot be crossed by the curve $\mu(T)$ because along it the self-energy would be infinite. 
The solution of the particle number equation makes then the chemical potential to osculate the curve $\mu_{\rm c}(T)$ when the temperature is lowered, and eventually to approach the limiting value $-\varepsilon_{0}/2$ when $T \to 0$.  
This behavior, albeit correct in the strong-coupling region, is unphysical for weak and intermediate couplings.  
Empirically, one can identify the temperature where $\mu(T)$ reaches its maximum as the temperature below which the proximity to this unphysical divergence begins to matter. 

We shall find below (see, in particular, Figs.~\ref{Figure-3} and \ref{Figure-4}) that the temperature range which is relevant to the pairing effects we are after extends well above this maximum at any given coupling.
Yet, to analyze in detail the single-particle spectral properties it is desirable to work at the lowest possible temperature, so as to reduce the importance of thermal broadening relative to interaction effects. 
As a consequence, we shall find it convenient to conduct this analysis at temperatures that correspond to (about) the maximum of $\mu(T)$ in Fig.~\ref{Figure-1}.

\vspace{-0.2cm}
\section{Comparison with the experi- \\ mental data for the trapped system} 
\label{sec: comparison-with-data}
\vspace{-0.2cm}

\begin{center}
{\bf A. Energy distribution curves}
\end{center}
\vspace{-0.2cm}

In the absence of final-state effects (as appropriate to the $^{40}\mathrm{K}$ atoms utilized in the experiment of Ref.~\onlinecite{Koehl-2011}), the radio-frequency spectral intensity is given by:
\begin{equation}
\mathrm{RF}(\tilde{\omega})=\frac{2}{N}\int \!\! d \mathbf{r} \int \!\! \frac{d \mathbf{k}}{(2 \pi)^{2}} A(\mathbf{k}, \xi(\mathbf{k};\mathbf{r}) -  \tilde{\omega})f_{F}(\xi(\mathbf{k};\mathbf{r}) -\tilde{\omega}).             
\label{total-RF-signal}
\end{equation}

\noindent
Here, $\tilde{\omega} = \omega_{\mathrm{rf}} - \omega_{\rm a}$ is the detuning frequency defined as the difference between the radio-frequency $\omega_{\rm rf}$ and the atomic 
transition frequency $\omega_{\rm a}$ for free atoms, $\mathbf{r}$ the position in the 2D trap, and 
$\xi(\mathbf{k};\mathbf{r}) = \mathbf{k}^{2}/(2m) - \mu({\bf r})$ where $\mu({\bf r})=\mu- V({\bf r})$ with the potential $V(\mathbf{r}) = \frac{1}{2} m  \omega_{0} \mathbf{r}^{2}$ trapping $N$ atoms.
The pre-factor in Eq.~(\ref{total-RF-signal}) is chosen to make the total area of the radio-frequency spectral intensity equal unity, namely, $\int_{-\infty}^{+\infty} d \tilde{\omega}\, {\rm RF}(\tilde{\omega})=1$. 
The spectral function $A(\mathbf{k}, \xi(\mathbf{k};\mathbf{r}) -  \tilde{\omega})$ is calculated  at position ${\bf r}$ in the trap within a local-density approximation.

The radio-frequency spectral intensity can be analyzed into its individual $\mathbf{k}$-components by exchanging the order of the two integrals in Eq.~(\ref{total-RF-signal}).   
One may thus define for each $\mathbf{k}$-component:
\vspace{-0.1cm}
\begin{equation}
\mathrm{RF}({\bf k},\tilde{\omega})=\frac{2}{N}\int \!\! d \mathbf{r}  A(\mathbf{k}, \xi(\mathbf{k};\mathbf{r}) -  \tilde{\omega})f_{F}(\xi(\mathbf{k};\mathbf{r}) -\tilde{\omega})             
\label{RFk}
\end{equation}

\vspace{-0.15cm}
\noindent
such that $\mathrm{RF}(\tilde{\omega})=\int\!\!\frac{d \mathbf{k}}{(2 \pi)^{2}} {\rm RF}({\bf k},\tilde{\omega})$.
By expressing energies in units of the trap Fermi energy $E_{F} = \omega_{0} N^{1/2}$, wave vectors in units of the Fermi wave vector $k_{F} = (2 m E_{F})^{1/2}$, and radial positions in units of the Thomas-Fermi radius $R_{F} = \sqrt{2 E_{F} / (m \omega_{0}^{2}) }$, one then gets the dimensionless expression: 
\vspace{-0.1cm}
\begin{equation}
\mathrm{RF}({\bf k},\tilde{\omega})=8 \int \!\! d \mathbf{r}  A(\mathbf{k}, \xi(\mathbf{k};\mathbf{r}) -  \tilde{\omega})f_{F}(\xi(\mathbf{k};\mathbf{r}) -\tilde{\omega}).             
\label{RFkbar}
\end{equation}
\vspace{-0.1cm}

Finally, to obtain an expression that can be directly compared with the experimental Energy Distribution Curve (EDC), it is sufficient to express the frequency $\tilde{\omega}$ in terms of the single-particle energy defined as $E_{s}= {\bf k}^2/(2m) - \tilde{\omega}$. 
This yields eventually:
\begin{equation}
{\rm EDC}({\bf k},E_s)=8 \int \!\! d \mathbf{r}  A(\mathbf{k}, E_s -\mu(\mathbf{r})) f_{F}(E_s -\mu(\mathbf{r})).             
\label{EDCkbar}
\end{equation}

\vspace{-0.3cm}
\begin{center}
{\bf B. Comparison with the experimental data}
\end{center}
\vspace{-0.1cm}

\begin{figure}[t]
\includegraphics[angle=0,width=7.2cm]{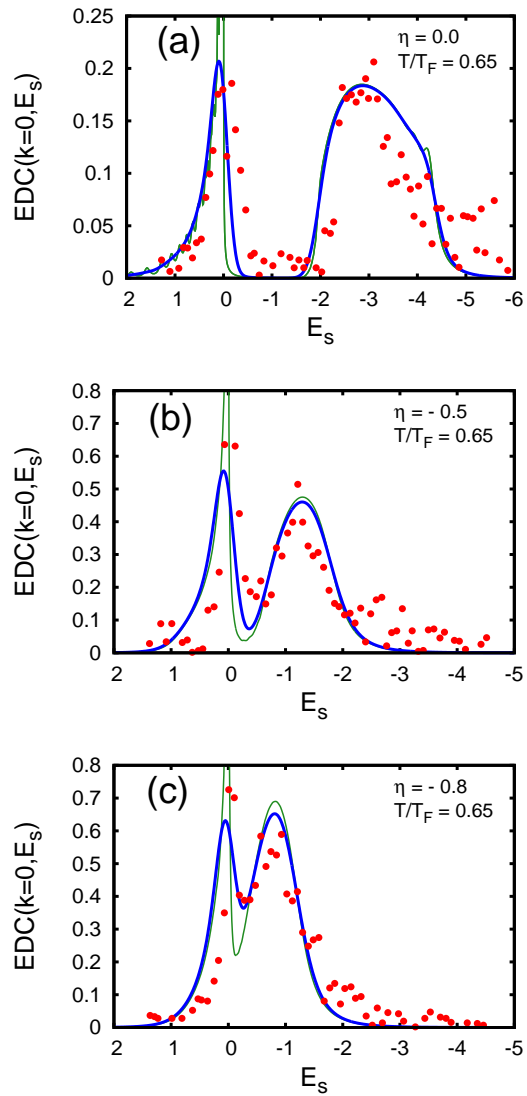}
\caption{(Color online) (a)-(c) The experimental spectra taken from the bottom panels of Fig.2(a) of Ref.~\onlinecite{Koehl-2011} (dots) are compared to our calculations of the
${\rm EDC}({\bf k}=0,E_s)$, without (thin lines) and with (thick lines) convolution with a Gaussian resolution of variance $1.5$ kHz.
In these panels, the temperature is fixed at $0.65 \, T_{F}$ where $T_{F}=E_{F}/k_{B}$, while the coupling parameter $\eta = - \ln(k_{F} a_{2D})$ takes different values. 
The single-particle energy $E_{s}$ is in units of $E_{F}$.}
\label{Figure-2}
\end{figure}

Figure \ref{Figure-2} compares the experimental data of Ref.~\onlinecite{Koehl-2011} for the energy distribution curve at ${\bf k}=0$ with our theoretical calculations that contain  
no fitting parameter. 
Specifically, for given coupling strength and temperature, the value of the chemical potential $\mu$ required to obtain EDC$({\bf k}= 0, E_s)$ is calculated by inverting numerically the number equation $N=\int d {\bf r} \, n({\bf r})$, where $ n({\bf r})$ is the particle number density obtained from the local spectral function at position ${\bf r}$. 
The experimental resolution of 1.5 kHz (as reported in the supplementary material of Ref.~\onlinecite{Koehl-2011}) is included in the comparison with the experimental data, by convoluting the theoretical results of Eq.~(\ref{EDCkbar}) with a normalized Gaussian of variance $\sigma=$1.5 kHz (once converted in units of $E_{F}$). 
For completeness, Fig.~\ref{Figure-2} reports the theoretical curves obtained both without (thin lines) and with (thick lines) this convolution.

The value of $E_{F}$ for the experimental data at  temperature $T/T_{F} = 0.65$ reported in Fig.~\ref{Figure-2} is determined to be  $11$ kHz.  
One should note that the value $k_F= 8.1 \mu$m$^{-1}$ given in Ref.~\onlinecite{Koehl-2011} would instead yield a Fermi energy $E_F= 8.4$ kHz. 
This value, however, refers to a lower temperature ($T/T_{F} = 0.27$) with less atoms in the sample due to evaporative cooling.  
The dependence of the number of particles on temperature in the experimental setup of Ref.~\onlinecite{Koehl-2011} was analyzed by the same group in Ref.~\onlinecite{Koehl-2012}. 
A linear interpolation of the Fermi wave vector data reported there for $T/T_{F} = 0.27$ and $T/T_{F} = 1.09$ (see the caption of their Fig.~3) allows us to conclude  that the value of the Fermi energy at  $T/T_{F} = 0.27$ has to be multiplied by a factor 1.3, in order to obtain the value of $E_{F}$ at $T/T_{F} = 0.65$. 
This yields the value $E_{F}= 11$ kHz quoted above, which fixes the horizontal scale of the experimental spectra. 
The vertical scale of the experimental spectra is instead fixed by making the height of the right experimental peak to coincide with the theoretical prediction. 
 
The comparison between the experimental data and our theoretical calculations shown in Fig.~\ref{Figure-2} appears to be rather good, to the extent that our calculations are able to reproduce not only the positions and widths of the experimental peaks and their evolution with coupling, but also the asymmetric shapes of the spectra.
These results give us confidence about the validity of our theoretical approach at least in the temperature range relevant to the experiment of Ref.~\onlinecite{Koehl-2011}. 
In this respect, it might be remarked that none of the previous theoretical works mentioned in the Introduction has attempted a direct comparison between the experimental spectra of  
Ref.~\onlinecite{Koehl-2011} and the corresponding theoretical calculations, in the way we have done in Fig.~\ref{Figure-2}. 

On the other hand, it has not been possible for us to compare our calculations also with the experimental data at $T/T_{F} = 0.27$ reported in the top panels of Fig.~2 of 
Ref.~\onlinecite{Koehl-2011}.   
This is because, for the couplings $\eta= (-0.8,-0.5,0.0)$ considered in Ref.~\onlinecite{Koehl-2011}, the non-self-consistent t-matrix approach yields the values 
$T_{c}/T_{F} = (0.37,0.45,0.58)$, respectively, for the the superfluid critical temperature $T_{c}$ of the 2D trapped system (cf.~Ref.~\onlinecite{Ohashi-2013}). 
The temperature $T/T_{F} = 0.27$ is then lower than our $T_{c}$, thus not allowing us to compare with the experimental data at this temperature (at least within the present theory formulated for the normal phase). 
In this context, a reference value for $T_{c}$ could be provided by the BEC limit $T_{c}/T_{F}=\sqrt{3}/\pi\simeq0.55$ that corresponds to an ideal Bose gas of molecules trapped in a two-dimensional harmonic  potential \cite{PS-2008}, because the superfluid critical temperature of the trapped two-dimensional Fermi gas should converge to this value when $\eta\to+\infty$.
Comparison with this value suggests that the values of $T_c$ quoted above for $\eta = (-0.8,-0.5,0.0)$, even though not expected {\it a priori\/} to be  quantitatively correct, are probably not unrealistic.

\begin{center}
{\bf C. Local couplings and normalized temperatures in the trap}
\end{center}
\vspace{-0.2cm}

Theoretically, the trap averaging required to compare with the experimental data rests on a local-density approximation, whereby the system is considered locally homogeneous 
with density $n(\mathbf{r})$.
To this end, once the density profile $n({\bf r})$ in the trap has been calculated for given values of the coupling parameter $\eta$ and temperature $T$, one can determine the {\em local\/} coupling parameter $\eta({\bf r})\equiv -\ln[k_F({\bf r}) a_{2D}]$ and normalized temperature $T/T_{F}({\bf r})$, where $k_{F}({\bf r})= \sqrt{2\pi n({\bf r})}$ and 
$k_{B} T_{F}({\bf r})=k_{F}({\bf r})^{2}/ (2m)$.

\vspace{-0.3cm}
\begin{table}[h]
\caption{Local values of coupling and normalized temperature for a 2D trapped Fermi gas at $T=0.65 T_{F}$.} \centering \begin{tabular}{c c c c c} \hline\hline
\phantom{a}$\eta$ \phantom{a}& \phantom{a}$\eta(0)\phantom{a}$ & \phantom{a}$T/T_F(0)$ \phantom{a}& \phantom{a}$\eta(r_{\rm max})$\phantom{a} &\phantom{a} $T/T_F(r_{\rm max})$\phantom{a}\\
\hline
-0.8 & -0.66 & 0.84 & -0.45 & 1.3 \\
-0.5 & -0.38 & 0.82 & -0.16 & 1.27 \\
 0.0  & -0.07 & 0.57 & 0.17 & 0.93 \\
\hline
\end{tabular}
\end{table}

Table I reports two examples of these local values for the three different cases considered in Fig.~\ref{Figure-2}, corresponding to the trap coupling parameter $\eta=(0.0,-0.5, -0.8)$ of
panels (a), (b), and (c), respectively, and $T=0.65 T_{F}$. 
Specifically, Table I reports the local values both at $r\equiv|{\bf r}|=0$ and at  $r=r_{\rm max}$ defined as the value of $r$ where $r  n(r)$ attains its maximum, such that the corresponding radial shell has the largest particle number.
The region $r\lesssim r_{\rm max}$ is, in fact, expected to contribute most to the measured EDC spectral intensity.

These local values of coupling and normalized temperature will be utilized in the next Section, to map the physical conditions that underlie the various panels of Fig.~\ref{Figure-2} onto the coupling vs temperature phase diagram of the homogeneous system.
Later on it will be also relevant to verity that \emph{locally} the system remains well above the critical temperature (cf. Fig.~\ref{Figure-7} below).
 
\vspace{-0.2cm}
\section{Properties of the underlying homogeneous system} 
\label{sec: homogeneous-system}
\vspace{-0.2cm}

\begin{center}
{\bf A. Boundary between the pseudo-gap and molecular regimes}
\end{center}
\vspace{-0.2cm}

\begin{figure}[h]
\includegraphics[angle=0,width=6.7cm]{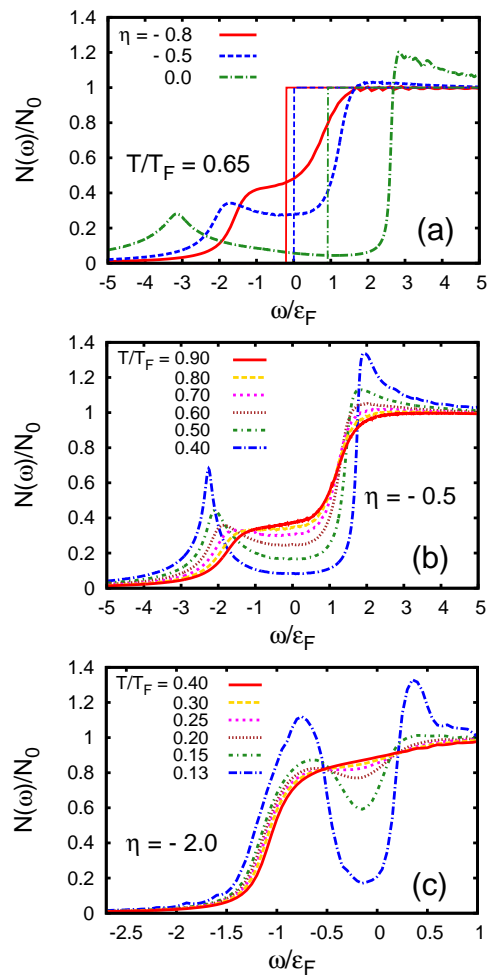}
\caption{(Color online) Single-particle density of states $N(\omega)$ of a homogeneous system (normalized to the non-interacting value $N_{0}=m/\pi$) vs the frequency $\omega$ (in units of the Fermi energy $\varepsilon_{F}$), for several values of coupling and temperature.
In panel (a) the function $N_{0}(\omega) = N_{0} \Theta(\omega + \mu)$ (where $\mu$ refers to the interacting system) is also shown for comparison.}
\label{Figure-3}
\end{figure}

We begin by considering the issue of the \emph{boundary between the pseudo-gap and molecular regimes}.
To this end, Fig.~\ref{Figure-3} shows the frequency dependence of the single-par-

\noindent
ticle density of states $N(\omega)$ obtained from Eq.(\ref{density-of-states}).
Specifically, panel (a) shows $N(\omega)$ for the same temperature and couplings of Fig.~\ref{Figure-2}, while panel (b) shows $N(\omega)$ for the coupling $\eta = -0.5$ and several temperatures about that of Fig.~\ref{Figure-2}.
From these curves one may identify the occurrence of a ``pseudo-gap'' whenever $N(\omega)$ has a local minimum about $\omega = 0$, such that the closing of this pseudo-gap occurs as soon as $N(\omega)$ becomes a monotonically increasing function of $\omega$.
In addition, panel (c) shows the temperature dependence of $N(\omega)$ for the weaker coupling $\eta = -2.0$, where a more symmetric shape of $N(\omega)$ emerges which resembles what is obtained in cuprate superconductors from tunneling experiments \cite{Ando-2007}. 

\begin{figure}[h]
\includegraphics[angle=0,width=7.5cm]{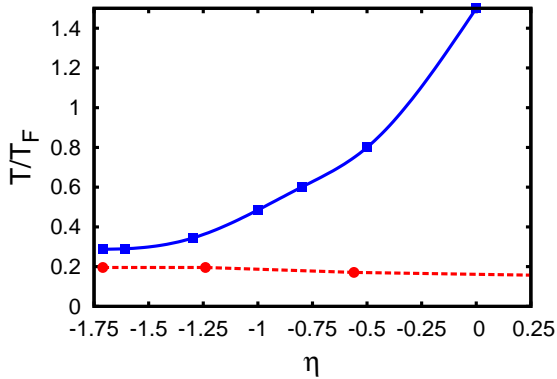}
\caption{(Color online) Coupling dependence of the pairing temperature $T^{*}$ where the pseudo-gap closes for a homogeneous 2D Fermi gas (squares and interpolating line).
The critical temperature $T_{c}$ determined experimentally in Ref.~\onlinecite{Jochim-2014} is also reported for comparison (circles and interpolating line).
Both temperatures are in units of $T_{F}=\varepsilon_{F}/k_{B}$.}
\label{Figure-4}
\end{figure}

Through this kind of analysis, we obtain the coupling dependence of the \emph{pairing temperature} $T^{*}$ at which the pseudo-gap closes according to the above criterion.
The result is reported in Fig.~\ref{Figure-4}, where for reference we report also the experimental data for the critical temperature $T_{c}$ in 2D that were recently determined in 
Ref.~\onlinecite{Jochim-2014}. 
This comparison shows that the pairing temperature $T^{*}$ at which the pseudo-gap closes in the single-particle density of states $N(\omega)$ becomes rapidly much larger than
$T_{c}$ as the coupling strength increases toward the BEC limit.

\begin{figure}[t]
\includegraphics[angle=0,width=8.5cm]{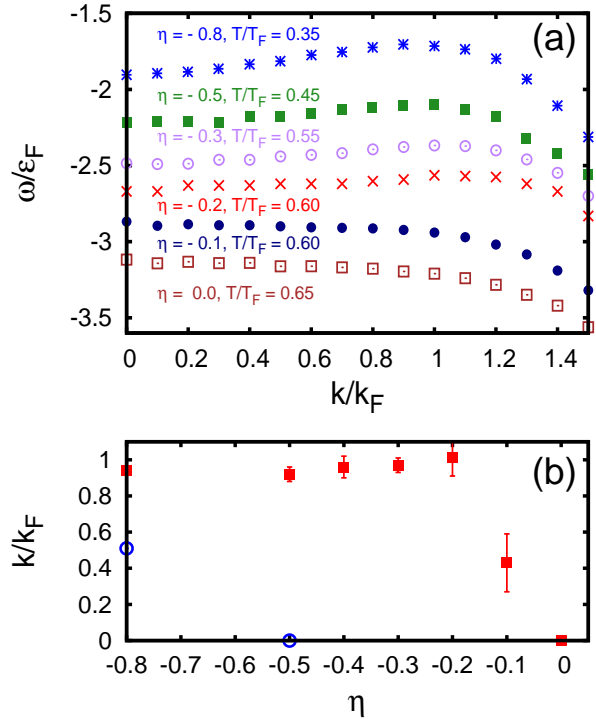}
\caption{(Color online) (a) Dispersion relations of the peak of the single-particle spectral function at negative frequencies, for various couplings and temperatures. 
The frequency $\omega$ is in units of $\varepsilon_{F}$ and the wave vector $k$ in units of $k_{F}$. 
(b) Coupling dependence of the Luttinger wave vector $k_{L}$ (squares with error bars) at which the back-bending occurs in the dispersions of panel (a) (each point corresponds to a different temperature as indicated in panel (a)). The coupling dependence of the wave vector $k_{\mu} = \sqrt{2 m \mu(T)}$ is also reported for comparison (empty circles).}
\label{Figure-5}
\end{figure}

What is not, however, evident from the above analysis is whether what has been identified as a ``pseudo-gap'' is a truly many-body effect or rather a manifestation of the two-body binding.
To answer this question, we perform here for the 2D system an analysis similar to what was done in Refs.~\onlinecite{Camerino_Jila-2011,PPPS-2012} for the 3D system.
Accordingly, in Fig.~\ref{Figure-5}(a) we show the dispersion relations obtained by following the evolution of the frequency position of the peak of the single-particle spectral function $A(\mathbf{k},\omega)$ at negative frequencies when the magnitude $k = |\mathbf{k}|$ of the wave vector is increased from $k=0$ to values larger than $k_{F}$, for several couplings and temperatures.
For each coupling, the temperature is chosen well below $T^{*}$ (that is, $T \approx 0.5 \, T^{*}$) yet above the temperature range where the chemical potential is influenced by the effects discussed in Fig.~\ref{Figure-1}.
From these dispersions one can identify the values of the Luttinger wave vector $k_{L}$ at which the curves ``back-bend'', thereby signaling the presence of an underlying Fermi surface.
The resulting coupling dependence of $k_{L}$ is shown in Fig.~\ref{Figure-5}(b), where the error bars correspond to the statistical error of BCS-like fits to the curves of panel (a) \cite{PPPS-2012}.
From this analysis we conclude that in 2D the boundary between the pseudo-gap and molecular regimes, where $k_{L}$ vanishes and the underlying Fermi surface disappears, lies in the range $-0.1 \lesssim \eta \lesssim 0.0$.
A corresponding analysis made in 3D had shown \cite{PPPS-2012} that $k_{L}$ vanishes at $T_{c}$ for the coupling $(k_{F} a_{3D})^{-1} \simeq 0.6$.

It is relevant to mention that an alternative (zero-temperature) criterion has sometimes been used in the literature to separate the BCS from the BEC regimes 
\cite{Leggett-1980,Randeria-1989,Sa-de-Melo-2000,Parish-2013}, by imposing the condition $\mu(T=0) = 0$ (which corresponds to where the back-bending of the single-particle 
dispersion occurs at $k=0$ at the level of the BCS mean field).
This condition can be utilized at any temperature $T$, and the coupling dependence of the wave vector $k_{\mu} = \sqrt{2 m \mu(T)}$ can correspondingly be identified 
for given $T$.
In general, the coupling dependence of $k_{\mu}$ cannot be expected to coincide with that of the Luttinger wave vector $k_{L}$, in terms of which we have identified the boundary between the pseudo-gap and the molecular regimes.
To show this difference for the specific problem in 2D, we have reported in Fig.~\ref{Figure-5}(b) also the coupling dependence of $k_{\mu}$ for the same temperatures at which $k_{L}$ 
was obtained. 
[A similar analysis in 3D was reported in Refs.~\onlinecite{Camerino_Jila-2011,PPPS-2012}.]

On physical grounds, the difference between $k_{L}$ and $k_{\mu}$ at given coupling and temperature is due to the effect on $k_{L}$ of a (diagonal) self-energy shift over and above the thermodynamic chemical potential $\mu$, shift which is then not present in $k_{\mu}$ by its very definition.
The only case when $k_{L}$ and $k_{\mu}$ coincide with each other is that of the BCS mean-field approach, where only an off-diagonal self-energy appears.
In particular, from Fig.~\ref{Figure-5}(b) one sees that $k_{\mu}$ vanishes in 2D at about $\eta = -0.5$ while $k_{L}$ vanishes at about $\eta = 0$.
This implies that the criterion $k_{\mu} = 0$ sets the BEC side of the crossover at $\eta \gtrsim -0.5$ as argued \cite{footnote} in Ref.~\onlinecite{Parish-2013}, while the criterion $k_{L} = 0$ adopted here sets the same side at $\eta \gtrsim 0.0$.
This difference is crucial when identifying the boundary between pseudo-gap and molecular regimes in an appropriate way.

\begin{figure}[t]
\includegraphics[angle=0,width=6.8cm]{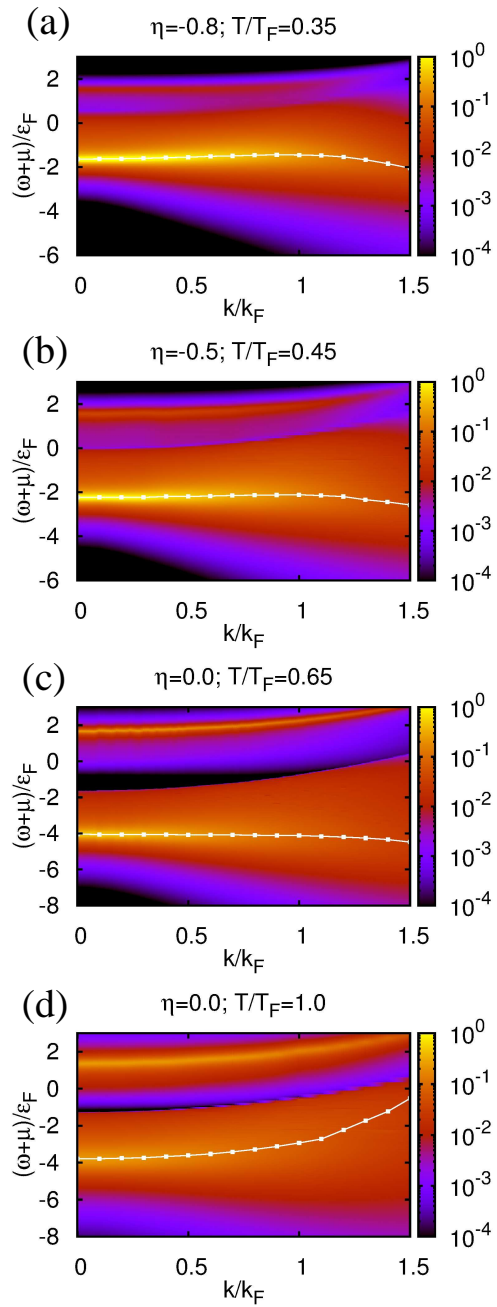}
\caption{(Color online) False color intensity plots of the single-particle spectral function $A(k,\omega)$ multiplied by the occupation factor $f_{F}(\omega)$, for several couplings and temperatures. 
The frequency $\omega$ is shifted by the chemical potential $\mu$ at given coupling and temperature (with $\mu$ taken from Fig.÷\ref{Figure-1}).}
\label{Figure-6}
\end{figure}

To get a deeper insight into the dispersion relations of Fig.~\ref{Figure-5}(a), we report in Fig.~\ref{Figure-6} false color intensity plots of the single-particle spectral function $A(k,\omega)$ multiplied by the occupation factor $f_{F}(\omega)$, for several couplings and temperatures. 
This is the quantity which underlies the average (\ref{EDCkbar}) for a trapped Fermi gas and is also measured in photoemission experiments in solid-state systems. 
In particular, panels (a)-(c) of Fig.~\ref{Figure-6} correspond to three cases considered in Fig.~\ref{Figure-5}(a) and reproduce the raw data from which the dispersions of 
Fig.~\ref{Figure-5}(a) were determined.
For completeness, we also report these dispersions in Fig.~\ref{Figure-6} (white circles and lines).
Note that for the frequencies relevant to these dispersions the occupation factor $f_{F}(\omega)$ is essentially equal to one. 

Panel (d) of Fig.~\ref{Figure-6}, on the other hand, corresponds to the case $\eta=0$ and $T=T_{\rm F}$.  
This panel can  thus be compared directly with the virial expansion results presented in Fig. 4(a) of Ref.~\onlinecite{Parish-2013} for the same coupling and temperature. 
This comparison shows that the virial expansion of Ref.~\onlinecite{Parish-2013} and the present t-matrix approach (which reduces to the virial expansion at high temperature \cite{Combescot-2006}) at this temperature still differ significantly from each other.  
It turns out, in particular, that the virial expansion misses a sizable self-energy shift in the quasi-particle dispersion at positive frequency and differs quantitatively from the t-matrix results also at negative frequency.  
Panel (d) of Fig.~\ref{Figure-6} can further be compared with Fig.~\ref{Figure-2}(a) since, according to Table I, the corresponding local values of coupling and normalized temperature 
of Fig.~\ref{Figure-2}(a) are close to those considered here.  
Specifically, a comparison can be made between the data of Fig.~\ref{Figure-6}(d) at $k=0$ and the trap average (\ref{EDCkbar}), having in mind that most of the signal of EDC$(0,E_s)$ comes from the region around $r_{\rm max}$.   
When making this comparison, one should recall that $(\omega +\mu)/\varepsilon_{F}$ in Fig.~\ref{Figure-6}(d) corresponds in Fig.~\ref{Figure-2} to $E_{s}$ multiplied by the factor 
$E_{F}/\varepsilon_{F}(r_{\rm max})=1.43$.
By taking into account this factor, one verifies that the position and width of the peak at negative energy in Fig.~\ref{Figure-2}(a) are consistent with the corresponding values extracted from the signal of Fig.~\ref{Figure-6}(d) at $k=0$.    
In addition, the peak at about zero energy in Fig.~\ref{Figure-2}(a) is essentially determined by the presence of free atoms in the outer shell of the trapped cloud, 
while the tail of the spectrum at $E_{s} \simeq 1$ (in units of $E_{F}$) in Fig.~\ref{Figure-2}(a) is contributed by the signal at positive frequency in Fig.~\ref{Figure-6}(d).  

This analysis shows that a comparison between the calculations for the homogenous system and the trap-averaged data of Ref.~\onlinecite{Koehl-2011} is meaningful only when supported by extended calculations for the trapped system, which allow for the determination of the relevant local couplings and temperatures and of the associated energy-conversion factors.  
For these reasons, when attempting such a comparison indirectly without the support of calculations for the trapped system (as it was done in Ref.~\onlinecite{Parish-2013}) one may end up with somewhat misleading results and conclusions about the relevance of the data.

\begin{figure}[t]
\includegraphics[angle=0,width=7.5cm]{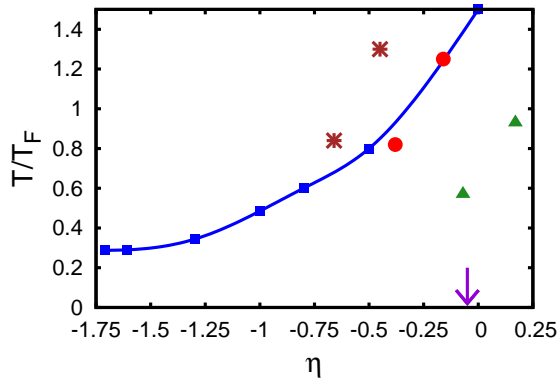}
\caption{(Color online) Local couplings and normalized temperatures of Table I are compared with the coupling dependence of $T^*$ (in units of  $T_{F}=\varepsilon_{F}/k_{B}$) 
of the homogeneous gas (squares and interpolating line). 
Asterisks correspond to the first row, circles to the second row, triangles to the third row of Table I, respectively. 
In each case, the values at $r=0$ lie on the left of the values at $r_{\rm max}$.
In addition, the vertical arrow indicates the position of the boundary between the pseudo-gap and molecular regimes determined in Fig.÷\ref{Figure-5}(b).}
\label{Figure-7}
\end{figure}

Finally, the local values of coupling and normalized temperature at $r=0$ and $r=r_{\rm max}$ that were reported in Table I can be compared with the coupling dependence of $T^*$ and with the boundary between the pseudo-gap and molecular regimes, so as to identify the coupling and temperature regions of the homogeneous gas which were explored in the experimental data of Fig.~\ref{Figure-2}.  
This comparison is made in Fig.~\ref{Figure-7}, from which one concludes that: 
(i) the data of Fig.~\ref{Figure-2}(a) (corresponding to the triangles in Fig.~\ref{Figure-7}) are below $T^*$ and lie at the boundary between the pseudo-gap and molecular regimes; 
(ii) the data of Fig.~\ref{Figure-2}(b) (corresponding to the circles in Fig.~\ref{Figure-7}) are just below $T^*$ and well within the pseudo-gap regime; 
(iii) the data of Fig.~\ref{Figure-2}(c) (corresponding to the asterisks in Fig.~\ref{Figure-7}) are above $T^*$ although at couplings consistent with the pseudo-gap regime at lower temperatures.
We thus conclude that the experiment of Ref.~\onlinecite{Koehl-2011} was able to explore also the pseudo-gap regime of most physical interest.

\vspace{-0.1cm}
\begin{center}
{\bf B. A unifying variable for 2D and 3D}
\end{center}
\vspace{-0.1cm}

One may take advantage of the similarities between pseudo-gap phenomena in 3D and 2D to \emph{identify a variable alternative to the coupling}, in terms of which it appears possible to unify the evolution from BCS to BEC in 3D and 2D.
Following Ref.~\onlinecite{PS-1994}, we identify this variable with the ratio between the pair size $\xi_{\mathrm{pair}}$ and the average inter-particle distance $d_{n}$ given by $[3/(4 \pi n)]^{1/3}$ in 3D and by $[1/(\pi n)]^{1/2}$ in 2D, where $\xi_{\mathrm{pair}}$ is obtained at $T=0$ within mean field in the two cases \cite{PS-1994,Randeria-1990}.

\begin{figure}[h]
\includegraphics[angle=0,width=7.0cm]{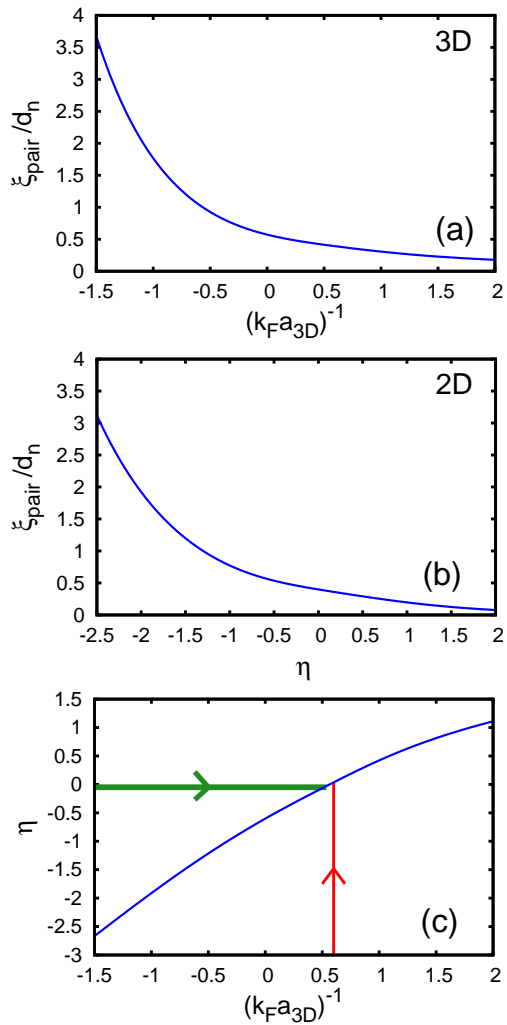}
\caption{(Color online) (a) Coupling dependence of the $T=0$ ratio between the pair size $\xi_{\mathrm{pair}}$ and the inter-particle spacing $d_{n}$ in 3D, and 
(b) corresponding dependence in 2D. 
(c) Relation between the couplings in 3D and 2D as obtained from the universal variable $\xi_{\mathrm{pair}}/d_{n}$. 
Segments with arrows identify (within numerical errors) the critical values of the couplings in 3D (thick line) and in 2D (thin line) where $k_{L}$ vanishes.}
\label{Figure-8}
\end{figure}

Figure \ref{Figure-8} shows the ratio $\xi_{\mathrm{pair}} / d_{n}$ vs the respective couplings, in panel (a) for 3D and in panel (b) for 2D.
From these curves, a relationship between the couplings $(k_{F} a_{3D})^{-1}$ in 3D and $\eta$ in 2D can be established.
The resulting curve is shown in Fig.~\ref{Figure-8}(c).
From these plots one verifies that the critical values of the couplings at which $k_{L}$ vanishes in 3D and 2D correspond to \emph{the same value} $(\simeq 0.4)$ of 
$\xi_{\mathrm{pair}} / d_{n}$.
This rather remarkable result justifies \emph{a posteriori} our identification of the ratio $\xi_{\mathrm{pair}} / d_{n}$ as the appropriate parameter that lies at the heart of the physics of the BCS-BEC crossover. 
From Fig.~\ref{Figure-8}(c) one further verifies that to the unitary limit $(k_{F} a_{3D})^{-1}=0$ in 3D there corresponds the value $\eta \simeq -0.6$ in 2D, for which 
$\xi_{\mathrm{pair}} / d_{n} \simeq 0.6$.

\begin{figure}[h]
\includegraphics[angle=0,width=7.5cm]{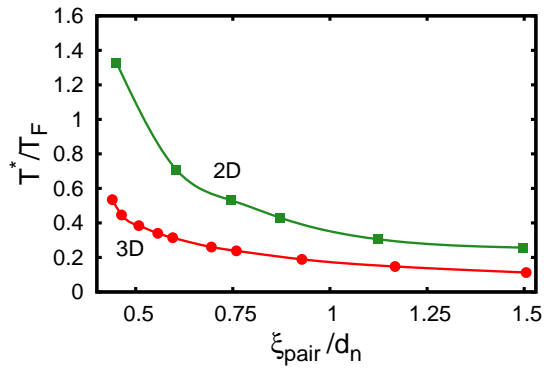}
\caption{(Color online) Temperature $T^{*}$ (in units of the Fermi temperature $T_{F}$) vs $\xi_{\mathrm{pair}}/d_{n}$ both in 3D (circles - from Ref.\onlinecite{Ohashi-2009}) 
and in 2D (squares - present calculation), where the lines interpolate through the calculated values.}
\label{Figure-9}
\end{figure}

In terms of this universal variable $\xi_{\mathrm{pair}} / d_{n}$, we can eventually reconsider the issue of the temperature $T^{*}$ up to which pseudo-gap phenomena survive.
To this end, Fig.~\ref{Figure-9} shows the dependence of $T^{*}$ on $\xi_{\mathrm{pair}} / d_{n}$ both for the 3D and 2D cases.
This direct comparison in terms of the same variable in both cases evidences in a quantitative way \emph{a strong enhancement of pseudo-gap effects} due to reduced dimensionality. 
For instance, to the ``unitary'' value $\xi_{\mathrm{pair}} / d_{n} \simeq 0.6$ there corresponds $(T^{*}/T_{F})_{3D} = 0.3$ and $(T^{*}/T_{F})_{2D} = 0.7$, with a $2.3$ amplification factor.

\vspace{-0.3cm}
\section{Concluding remarks}
\label{sec:conclusions}
\vspace{-0.3cm}

In this paper, we have dealt with the problem of the boundary between the pseudo-gap and molecular regimes for a 2D Fermi gas with attractive inter-particle interaction, a problem that can also be rephrased in the context of the way in which many-body effects hinge on two-body effects.
Specifically in 2D, this appears to be a rather delicate question since a two-body (molecular) bound state exists for all values of the attractive fermionic interaction.
To answer this question, we have presented a theoretical approach that has proven capable to reproduce quantitatively the experimental spectra of 
Ref.~\onlinecite{Koehl-2011}, assessing in this way that that experiment was able to explore also the pseudo-gap regime of most interest.

This turns out to be a particularly important result, to the extent that in 3D, on the other hand, the experimental detection of the pseudo-gap regime above $T_{c}$ at unitarity has been widely debated in the literature \cite{Camerino_Jila-2011,Salomon-2011}, in this context a general consensus having been achieved only on the value of the coupling (on the BEC side of unitarity) where the pseudo-gap fades away and the molecular regime sets in.

Also for this reason it was relevant to provide a unifying description of the pseudo-gap regime for the 2D and 3D cases.
This was achieved by spanning the BCS-BEC crossover through a ``universal'' variable, which is identified by a common definition in 2D and 3D and supersedes the most often used coupling variable that depends instead on the two-body scattering specific to dimensionality.
In terms of this new variable, it has been possible to reckon what can be regarded as the analog of the unitarity limit for a Fermi system in 2D, and further to establish that the temperature range where the pseudo-gap survives in 2D is quite enlarged with respect to 3D.

All these considerations, about the issue of the pseudo-gap due to pairing correlations in 2D, appear to be relevant also in the context of high-temperature (cuprate) superconductors where this issue has been highly debated over the last several years.
This is because of the presence in these systems of other effects that may concur with pairing in the formation of a pseudo-gap in the single-particle spectra \cite{Kaminski-2011,Kapitulnik-2008,Kaminski-2014}.
What we have explicitly shown here, that in 2D a pseudo-gap due to pairing persists in the normal phase over a quite wide temperature range just when the pair size is comparable to the average inter-particle spacing, suggests, in fact, that pairing correlations cannot be dismissed when addressing the pseudo-gap issue in high-temperature superconductors.


\begin{center}
\begin{small}
{\bf ACKNOWLEDGMENTS}
\end{small}
\end{center}
\vspace{-0.2cm}

F. Marsiglio acknowledges partial support from the University of Camerino during his sabbatical year.

\vspace{-0.4cm}                                                                                                                                                                                                                                                                                                                                                                                                         
\appendix
\section{BEC LIMIT FOR COMPOSITE BOSONS IN 2D WITHIN SELF-CONSISTENT \\ AND NON-SELF-CONSISTENT $\mathrm{t}$-MATRICES}
\label{sec:appendix-A}
\vspace{-0.3cm} 

In 2D, the pair propagator (\ref{pair-propagator}) takes the following asymptotic form in the BEC limit when $\mu/\varepsilon_{F} \rightarrow - \infty$:
\vspace{-0.25cm}
\begin{equation}
\Gamma_{0}(\mathbf{q},\Omega_{\nu}) \simeq - \, \frac{4 \pi \varepsilon_{0}}{m} \, \frac{1}{i \Omega_{\nu} + \mu_{B} - \mathbf{q}^{2}/(4m)}
\label{pair-propagator-BEC-limit}
\end{equation}

\vspace{-0.25cm}
\noindent
where $\mu_{B} = 2 \mu + \varepsilon_{0}$ is the chemical potential of composite bosons which form out of fermion pairs in this limit.
Apart from an overall constant factor, the form (\ref{pair-propagator-BEC-limit}) corresponds to the Green's function of non-interacting composite bosons.
As a consequence, no interaction among composite bosons survives in this limit within the non-self-consistent t-matrix approach, since only one 
$\Gamma_{0}$ appears in the fermionic single-particle self-energy (\ref{self-energy}).

The situation is somewhat different within the self-consistent t-matrix approach, where the dressed Green's function $G$ replaces the bare $G_{0}$
in the bubble (\ref{regularized-pp-bubble}).
In the BEC limit, this replacement has the result of introducing an effective interaction among composite bosons, as it can be seen, e.g., from Fig.4 of Ref.~\onlinecite{PS-2000}.
Referring, in addition, to Fig.3(b) of the same Ref.~\onlinecite{PS-2000}, in the limit of vanishing wave vectors and frequencies the value of this effective interaction in 2D turns out to be $4 \pi/ m$ independent of the 2D scattering length $a_{2D}$, while in 3D this interaction equals $4 \pi a_{3D}/ m$ where $a_{3D}$ is the 3D scattering length.
This implies that, within the self-consistent t-matrix approach, an effective interaction among composite bosons survives even in the extreme BEC limit, a result which 
contradicts one's physical intuition about the diluteness of the Bose gas.

As a matter of fact, the exact solution for the scattering problem of two dimers in vacuum given in Ref.~\onlinecite{Petrov-2003} yields for the dimer-dimer scattering length the value $a^{B}_{2D} = 0.55 \, a_{2D}$, which vanishes with $a_{2D}$ in the extreme BEC limit. 
A proper summation of diagrams in the limit of zero density would thus be required  beyond the self-consistent t-matrix to reproduce this result in 2D, along the lines of the approach used in Ref.~\onlinecite{Kagan-Combescot-2006} for the 3D case.




\begin{thebibliography}{99}

\bibitem{Koehl-2011}
M. Feld, B. Fr\"{o}hlich, E. Vogt, M. Koschorreck, and M. K\"{o}hl, Nature {\bf 480}, 75 (2011).

\bibitem{Koehl-2012}
B. Fr\"{o}hlich, M. Feld, E. Vogt, M. Koschorreck, M. K\"{o}hl, C. Berthod, and T. Giamarchi, Phys. Rev. Lett. {\bf 109}, 130403 (2012).

\bibitem{Sharapov-2001}
V. M. Loktev, R. M. Quick, and S. Sharapov, Phys. Rep. {\bf 349}, 1 (2001).

\bibitem{Larkin-2005}
A. Larkin and A. A. Varlamov, {\em Theory of Fluctuations in Superconductors} (Oxford University Press, Oxford, 2005).

\bibitem{Campuzano-2008}
A. Kanigel, U. Chatterjee, M. Randeria, M. R. Norman, G. Koren, K. Kadowaki, and J. C. Campuzano, Phys. Rev. Lett. {\bf 101}, 137002 (2008).

\bibitem{Kaminski-2009}
T. Kondo, R. Khasanov, T. Takeuchi, J. Schmalian, and A. Kaminski, Nature {\bf 457}, 296 (2009).

\bibitem{Shen-2010}
M. Hashimoto, R.-H. He, K. Tanaka, J.-P. Testaud, W. Meevasana, R. G. Moore, D. Lu, H. Yao, Y. Yoshida, H. Eisaki, T. P. Devereaux,  Z. Hussain, and Z.-X. Shen, Nat. Phys. {\bf 6}, 414 (2010).

\bibitem{Kaminski-2011}
T. Kondo, Y. Hamaya, A. D. Palczewski, T. Takeuchi, J. S. Wen, Z. J. Xu, G. Gu, J. Schmalian, and A. Kaminski, Nat. Phys. {\bf 7}, 21 (2011).

\bibitem{Lanzara-2013}
W. Zhang, C. L. Smallwood, C. Jozwiak, T. L. Miller, Y. Yoshida, H. Eisaki, D.-H. Lee, and A. Lanzara, Phys. Rev. B {\bf 88}, 245132 (2013).

\bibitem{Jin-2008} 
J. T. Stewart, J. P. Gaebler, and D. S. Jin, Nature {\bf 454}, 744 (2008).
	
\bibitem{Jila_Camerino-2010}
J. P. Gaebler, J. T. Stewart, T. E. Drake, D. S. Jin, A. Perali, P. Pieri, and G. C. Strinati, Nat. Phys. {\bf 6}, 569 (2010).

\bibitem{PS-1994}  
F. Pistolesi and G. C. Strinati, Phys. Rev. B {\bf 49}, 6356 (1994).

\bibitem{Pietila-2012} 
V. Pietil\"{a}, Phys. Rev. A {\bf 86}, 023608 (2012).

\bibitem{Ohashi-2013}
R. Watanabe, S. Tsuchiya, and Y. Ohashi, Phys. Rev. A {\bf 88}, 013637 (2013). 

\bibitem{Parish-2013}
V. Ngampruetikorn, J. Levinsen, and M. M. Parish, Phys. Rev. Lett. {\bf 111}, 265301 (2013).

\bibitem{Hofmann-2013}
M. Barth and J. Hofmann, Phys. Rev. A {\bf 89}, 013614 (2014).

\bibitem{Enss-2013} 
M. Bauer, M. M. Parish, and T. Enss, Phys. Rev. Lett. {\bf 112}, 135302 (2014).

\bibitem{Jochim-2014} 
M. G. Ries, A. N. Wenz, G. Zurn, L. Bayha, I. Boettcher, D. Kedar, P. A. Murthy, M. Neidig, T. Lompe, and S. Jochim, arXiv:1409.5373.

\bibitem{Combescot-2006} 
R. Combescot, X. Leyronas, and M. Y. Kagan, Phys. Rev. A {\bf 73}, 023618 (2006).

\bibitem{Camerino_Jila-2011} 
A. Perali, F. Palestini, P. Pieri, G. C. Strinati, J. T. Stewart,  J. P. Gaebler, T. E. Drake, and D. S. Jin, Phys. Rev. Lett. {\bf 106}, 060402 (2011).

\bibitem{PPSC-2002}
A. Perali, P. Pieri, G. C. Strinati, and C. Castellani, Phys. Rev. B {\bf 66}, 024510 (2002).

\bibitem{FW-1971} 
A. L. Fetter and J. D. Walecka, \emph{Quantum Theory of Many-particle Systems} (McGraw-Hill, New York, 1971).

\bibitem{Giorgini-2011}
G. Bertaina and S. Giorgini, Phys. Rev. Lett. {\bf 106}, 110403 (2011).

\bibitem{Schmitt-Rink-1989} 
S. Schmitt-Rink, C. M. Varma, and A. E. Ruckenstein, Phys. Rev. Lett. {\bf 63}, 445 (1989).

\bibitem{NSR-1985}
P. Nozi\`{e}res and S. Schmitt-Rink, J. Low. Temp. Phys. {\bf 59}, 195 (1985).

\bibitem{PS-2008}
C. J. Pethick and H. Smith, \emph{Bose-Einstein Condensation in Dilute Gases} (Cambridge University Press, Cambridge, 2008).

\bibitem{Ando-2007} 
K. K. Gomes, A. N. Pasupathy, A. Pushp, S. Ono, Y. Ando, and A. Yazdani, Nature {\bf 447}, 569 (2007).

\bibitem{PPPS-2012}
F. Palestini, A. Perali, P. Pieri, and G. C. Strinati, Phys. Rev. B {\bf 85}, 024517 (2012).

\bibitem{Leggett-1980}
A. J. Leggett, in \emph{Modern Trends in the Theory of Condensed Matter} (Springer-Verlag, Berlin, 1980), A. Pekalski and R. Przystawa, Eds., Lecture Notes in Physics, vol. 115, p. 13.

\bibitem{Randeria-1989}
M. Randeria, J-M. Duan, and L-Y. Shieh,  Phys. Rev. Lett. {\bf 62}, 2887 (1989).

\bibitem{Sa-de-Melo-2000}
R. D. Duncan and C. A. R. S\'{a} de Melo, Phys. Rev. B {\bf 62}, 9675 (2000).

\bibitem{footnote} 
In Ref.~\onlinecite{Parish-2013}, to set the boundary between the BCS and BEC regimes in 2D, the value of the chemical potential $\mu$ at zero temperature was taken from the QMC data of Ref.~\onlinecite{Giorgini-2011}. Since the QMC calculation goes beyond the BCS mean field, the identification of $k_{\mu}$ with $k_{L}$ as the place where the back-bending occurs is not appropriate.

\bibitem{Randeria-1990}
M. Randeria, J. M. Duan, and L. Y.  Shieh, Phys. Rev. B {\bf 41}, 327 (1990).

\bibitem{Ohashi-2009}
S. Tsuchiya, R. Watanabe, and Y. Ohashi, Phys. Rev. A {\bf 80}, 033613 (2009). 

\bibitem{Salomon-2011} 
S. Nascimb\`{e}ne, N. Navon, S. Pilati, F. Chevy, S. Giorgini, A. Georges, and C. Salomon, Phys. Rev. Lett. {\bf 106}, 215303 (2011).

\bibitem{Kapitulnik-2008}
J. Xia, E. Schemm, G. Deutscher, S. A. Kivelson, D. A. Bonn, W. N. Hardy, R. Liang, W. Siemons, G. Koster, M. M. Fejer, and A. Kapitulnik, Phys. Rev. Lett. {\bf 100}, 127002 (2008).

\bibitem{Kaminski-2014}
A. Kaminski, T. Kondo, T. Takeuchi, and G. Gu, arXiv:1403.0492.

\bibitem{PS-2000}
P. Pieri and G. C. Strinati, Phys. Rev. B {\bf 61}, 15370 (2000).

\bibitem{Petrov-2003}
D. S. Petrov, M. A. Baranov, and G. V. Shlyapnikov, Phys. Rev. A {\bf 67}, 031601(R) (2003).

\bibitem{Kagan-Combescot-2006}
I. V. Brodsky, M. Yu. Kagan, A. V. Klaptsov, R. Combescot, and X. Leyronas, Phys. Rev. A {\bf 73}, 032724 (2006). 

\end{thebibliography}
\end{document}